\documentclass[prb,preprint,showpacs,preprintnumbers,sortedaddress,amsmath,amssymb]{revtex4}
\usepackage[T1]{fontenc}   
\usepackage{graphicx}
\usepackage{dcolumn}
\usepackage{bm}

\usepackage{color}

\definecolor{Blue}{rgb}{0,0,1}
\definecolor{NavyBlue}{rgb}{0.14,0.14,0.56}
\definecolor{rot}{cmyk}{0,1,1,0}

\begin{document}

\title{Vibrational properties of graphene nanoribbons by first-principles calculations}
\author{Roland Gillen}
\author{Marcel Mohr}
\author{Janina Maultzsch}
\author{Christian Thomsen}
\affiliation{Institut für Festkörperphysik, Technische Universität Berlin, Hardenbergstr. 36, 10623 Berlin}

\date{\today}

\begin{abstract}
We investigated the vibrational properties of graphene nanoribbons by means of first-principles calculations on the basis of density functional theory.
We confirm that the phonon modes of graphene nanoribbons with armchair and zigzag type edges can be interpreted as fundamental oscillations and their overtones. These show a characteristic dependence on the nanoribbon width. Furthermore, we demonstrate that a mapping of the calculated $\Gamma$-point 
phonon frequencies of nanoribbons onto the phonon dispersion of graphene corresponds to an ``unfolding'' of nanoribbons' Brillouin zone onto that of graphene. We consider the influence of spin states with respect to the phonon spectra of zigzag nanoribbons and provide comparisons of our results with past studies.
\end{abstract}

\pacs{ 61.48.De,
63.22.-m, 63.20.D-
63.20.dk }
\maketitle

\section{Introduction}
The outstanding properties of graphene and graphene-related structures of nanosize gave rise to extensive theoretical and experimental research during the last two decades.
Along with the heavily studied carbon nanotubes (CNT) another quasi 1D-nanostructure aroused special interest:  Terminated stripes of graphene, so called graphene nanoribbons (GNR). Recent progress in preparation of single layered graphene sheets\cite{novoselov04,zhang05,berger06,novoselov07} allows the fabrication of GNRs through lithographic techniques\cite{han206805,Li08} and possibly the verification of theoretical predictions regarding electronic and optical properties.
In the course of such investigations interesting magnetic properties\cite{fujita96,wakabayashi98,wakabayashi8271,kusakabe092406,yamashiro193410,sonmetall,lee174431}, quasi-relativistic behavior of electrons and the possibility of bandgap engineering\cite{han206805,ezawa045432,barone606,son216803} by varying ribbon widths were shown. 
These results make GNRs seem promising for future developments in nanotechnology and nanoelectronics.

The propagation of valence electrons in graphene structures is accompanied by exceptionally strong electron-phonon coupling\cite{pisanibornoppen}. The investigation of the vibrational spectum in these materials is thus of fundamental importance for the electron transport in electronic devices and of great general interest.

In this paper, we present our studies of the $\Gamma$-point phonons of different armchair and zig-zag nanoribbons, obtained through \emph{ab-initio} density functional theory calculations. We found that it is possible to classify the $\Gamma$-point phonon modes of hydrogen passivated GNRs into fundamental oscillations, overtones and C-H vibrational modes. Fundamental oscillations and overtones can be mapped onto the graphene phonon dispersion by unfolding the GNR Brillouin zone onto that of graphene. Furthermore, we discuss the dependence of GNR phonon frequencies on the nanoribbon width. 

\begin{figure}
\includegraphics[width=1.1\columnwidth]{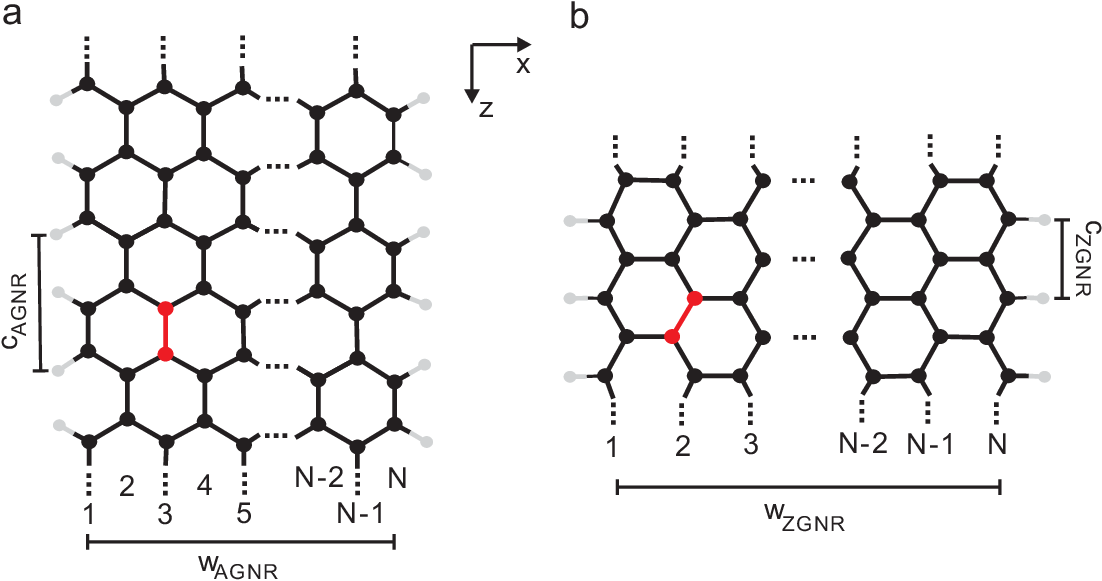}
\caption{\label{fig:structure} (color online) Structure of (a) a $N$-AGNR and (b) a $N$-ZGNR. In each case one dimer is emphasized in light grey (red). We found a lattice constant of $a_0=2.4656$ \AA\space for a relaxed sheet of graphene. The ideal lattice constants of the nanoribbons are then $c_{\mbox{\tiny AGNR}}=\sqrt{3}a_0=4.27$ \AA\space and $c_{\mbox{\tiny ZGNR}}=a_0=2.4656$ \AA. The corresponding ribbon widths, i.e. the distance between C atoms at opposing edges, are $w_{\mbox{$\mbox{\tiny AGNR}$}}=\frac{1}{2}(N-1)a_0$ and $w_{\mbox{$\mbox{\tiny ZGNR}$}}=\frac{\sqrt{3}}{2}(N-1)a_0$. The relaxation results in nanoribbon widths slightly below (for AGNRs) or above (for ZGNRs) the values calculated with these equations (see Tables ~\ref{tab:table1} and ~\ref{tab:table2}). The deviations however decrease with increasing ribbon width.}
\end{figure}

\section{Calculations}
Graphene nanoribbons can, at least in some cases, be understood as cut
and unrolled carbon nanotubes. These geometric similarities suggest
that the phonon spectra of comparable CNTs and GNRs may be similar. On
the other hand, unlike CNTs, nanoribbons possess edges, which have a lower
coordination number, and technically, they require special
treatment. A widespread method to take care of the carbon atoms at the edges in
calculations is to passivate the edges with atoms or molecules,
mainly hydrogen.  Because the diversity in the possibilities
to cut out GNRs of a graphene sheet is larger than to ``roll'' it into a seamless cylinder, i.e carbon nanotubes, the number of edge types is very large. This leads to the need for a
classification of those graphene nanoribbons.  Some approaches use a
$(p,q)$\cite{ezawa045432} type classification with two characteristic integers, similar
to the common classification of CNTs, or a classification that is based on the number of honeycombs along the ribbon width\cite{yamada08}.

For the purpose of this paper entirely sufficient is the approach of
Son et al. \cite{sonmetall}, where GNR are classified by their edge
type into armchair graphene nanoribbons (AGNR) and zig-zag nanoribbons
(ZGNR), and giving the number of dimers $N$ in the unit cell (see
Fig.~\ref{fig:structure}). The unit cell with $N$ dimers is extended
periodically along the $z$-direction, resulting in an infinitely long strip of graphene. 

We define the width of a graphene nanoribbon as the distance between the central points of the outmost dimers (refer to Fig.~\ref{fig:structure}). 
The ideal ribbon width, i.e. the width of an unrelaxed nanoribbon, is dependent on $N$ and given by
\begin{eqnarray}
w_{\mbox{\tiny AGNR}}&=&\frac{1}{2}(N-1)a_0\label{G4}\\
\text{and}\nonumber\\ 
w_{\mbox{\tiny ZGNR}}&=&\frac{\sqrt{3}}{2}(N-1)a_0\label{G5}
\end{eqnarray} 
with the graphene lattice constant $a_0$. In our calculations, the relaxed lattice constant is $a_0 = 2.4656$\AA.
The width of $N$-ZGNRs with odd $N$ is equivalent to the circumference of an $(\frac{N-1}{2},\frac{N-1}{2})$-CNT. Thus, these nanoribbons could be rolled into armchair nanotubes. For $N$-ZGNR with even $N$, however, there are no corresponding nanotubes. Similarly, the width of $N$-AGNRs with odd $N$ is equivalent to the chiral vector of a $(\frac{N-1}{2},0)$-CNT, whereas AGNRs with even $N$ do not correspond to any nanotubes.\\

We used density functional theory in the local approximation form
\cite{perdew81} to calculate $N$-AGNRs and $M$-ZGNRs with the number
of dimers per unit cell $N$=7..15 and $M$=4..14 respectively. 
Pseudopotentials were generated with the
Troullier-Martins scheme\cite{troullier91} for the following
valence-state configurations: C $2s^2(1.49), 2p^2(1.50)$; H
$1s^1(1.25)$, where the value in parenthesis indicates the
pseudopotential core radii in bohr. The valence electrons were
described by a double-$\zeta$ basis set plus an additional polarizing
orbital. The localization of the basis followed the standard split
scheme and was controlled by an internal {\sc SIESTA}\cite{siesta1,siesta2} parameter, the energy shift, for which
a value of 50\,meV was used. This resulted in basis functions with a
maximal extension of 3.31\,\AA\space (C) and 3.2\,\AA\space (H).
As {\sc SIESTA} works with periodic boundary conditions, the lattice vectors in direction perpendicular to the nanoribbon axis were scaled in such a way that the space between periodic images of the nanoribbons was at least
20\,\space\AA\space in order to prevent interaction between them. Real space integrations were performed on a grid with
a fineness of 0.08\,\AA, which can represent plane waves up to
an energy of 270\,Ry.

\begin{table}
\caption{\label{tab:table1}: Lattice constants $c$ and widths $w$ of various relaxed zigzag-edged graphene nanoribbons. $\Delta=\frac{c-c_{\mbox{\tiny ideal}}}{c_{\mbox{\tiny ideal}}}$ is the relative deviation of the calculated lattice constant from the ideal lattice constant.}
\begin{ruledtabular}
\begin{tabular}{ccccc}
\small$N$&$c$ (\AA)&$\Delta$\footnote{$c_{\mbox{\tiny ideal}}=2.4656$\space\AA}&$w$ (\AA)&$w_{\mbox{\tiny ideal}}$ (\AA)\footnote{calculated by $w_{\mbox{\tiny ideal}}=\frac{\sqrt{3}}{2}(N-1)a_0$}\\
\hline
2&2.461877&-0.15\%&2.141&2.135\\
3&2.461083&-0.18\%&4.286&4.27\\
4&2.461787&-0.15\%&6.427&6.406\\
5&2.462023&-0.14\%&8.566&8.541\\
6&2.46277&-0.11\%&10.702&10.676\\
7&2.465000&-0.02\%&12.838&12.811\\
8&2.46414&-0.06\%&14.975&14.947\\
9&2.464095&-0.06\%&17.113&17.082\\
10&2.464095&-0.06\%&19.249&19.217\\
12&2.464195&-0.06\%&23.523&23.488\\
14&2.464649&-0.04\%&27.794&27.758\\
16&2.464730&-0.04\%&32.067&32.029\\
\end{tabular}
\end{ruledtabular}
\end{table}

\begin{table}
\caption{\label{tab:table2}: Lattice constants $c$ and widths $w$ of various relaxed armchair-edged graphene nanoribbons.$\Delta=\frac{c-c_{\mbox{\tiny ideal}}}{c_{\mbox{\tiny ideal}}}$.}
\begin{ruledtabular}
\begin{tabular}{ccccc}
$N$&$c$ (\AA)& $\Delta$ \footnote{$c_{\mbox{\tiny ideal}}=\sqrt{3}a_0=4.27$ \space\AA}& $w$ (\AA)& $w_{\mbox{\tiny ideal}}$ (\AA)\footnote{calculated by $w_{\mbox{\tiny ideal}}=\frac{1}{2}(N-1)a_0$}\\
\hline
4&4.318&1.1\%&3.653&3.698\\
5&4.309&0.9\%&4.885&4.931\\
6&4.308&0.9\%&6.107&6.164\\
7&4.296&0.6\%&7.353&7.3968\\
8&4.293&0.5\%&8.587&8.629\\
9&4.294&0.5\%&9.809&9.862\\
10&4.289&0.4\%&11.056&11.095\\
11&4.288&0.4\%&12.185&12.328\\
12&4.288&0.4\%&13.511&13.561\\
13&4.287&0.4\%&14.753&14.794\\
14&4.284&0.3\%&15.988&16.026\\
15&4.284&0.3\%&17.212&17.259\\
16&4.284&0.3\%&18.454&18.492\\
17&4.284&0.3\%&19.689&19.725\\
20&4.28&0.2\%&23.381&23.423\\
21&4.28&0.2\%&24.609&24.656\\
22&4.28&0.2\%&25.846&25.889\\
\end{tabular}
\end{ruledtabular}
\end{table}
\normalsize

A minimum of 30 $k$-points equally spaced along the 1D Brillouin zone
was used.  The phonon calculations were performed with the method of
finite differences\cite{Yin82}. We fully relaxed the atomic positions
of both AGNRs and ZGNRs until the atomic forces of each atom were less
than 0.01 eV/\AA\space and minimized the total energy as function of lattice constant (refer to Tables ~\ref{tab:table1} and ~\ref{tab:table2}). 
We used a supercell approach with a $9\times9\times0$ supercell and the above paramaters to calculate the phonon dispersion of  graphene. These calculations resulted in a $\Gamma$-point frequency for the $E_{2g}$ modes of 1622 cm$^{-1}$.
This is slightly higher than the experimentally obtained graphene $E_{2g}$ frequency of about 1580 cm$^{-1}$\onlinecite{ferrarigraphene}. All calculated frequencies are therefore scaled by a constant $C$=0.974 to
achieve better comparability with experimental results.

\section{Results and discussion}
Graphene nanoribbons have a large length to width ratio, which results in a quasi-1D crystal-like behavior and
is expected to lead to confinement effects for the $\pi$ orbital electrons
perpendicular to the ribbon axis. It is therefore justified to regard the nanoribbons as infinitely long in our phonon calculations. Thus, the phonon wave vector in direction of the ribbon axis, $\bold
k_{\parallel}$, is quasi-continuous. The ribbon edges however only
allow standing waves perpendicular to the ribbon axis, and thus induce the boundary
condition
\begin{equation*}
\bold k_{\perp,n}\cdot w_{\mbox{\tiny ribbon}} = n\cdot \pi\\
\end{equation*}
on the phonon wave $f(r,t)=A\cdot e^{\bold k\cdot \bold r-\omega t}$, leading to a quantized wave vector 
\begin{equation}
\bold k_{\perp,n}=\frac{\pi}{w_{\mbox{\tiny ribbon}}}\cdot n\label{G2}\\
\end{equation} 
with the order of vibration $n=0..N-1$.

We expect therefore a vibrational behavior similar to that
of an elastic sheet or a chain of $N$ atoms with fixed or free ends, i.e., the
appearance of fundamental vibrations and overtones.  The phonon
spectrum of an $N$-AGNR or $N$-ZGNR should comprise of six fundamental
modes and 3$\cdot$2$N$-6 = 6($N$-1) overtones.
Therefore, in a given phonon spectrum, we should be able to assign $N$-1 overtone modes to each fundamental mode.\\

\subsection{Armchair Nanoribbons}
Our calculations yield for each AGNR a $\Gamma$-point phonon
spectrum consisting of 3$m$ modes, with $m$ = number of atoms per unit
cell. The atomic displacements of these $\Gamma$-point modes can be
classified into pure longitudinal (L), transverse (T) or out-of-plane
(Z) modes. Each $\Gamma$-point phonon mode can be associated with one of three types which will be discussed separately: (1)  C-H modes resulting from the passivation with hydrogen, (2) fundamental modes, or (3) overtones.

\subsubsection{C-H modes}
The four hydrogen atoms in the unit cell of AGNRs give twelve vibrational modes. These modes show large amplitudes of the hydrogen atoms in contrast to the almost negligibly small displacements of the carbon atoms. They can be grouped into 6 pairs of degenerate modes. We find C-H modes of different polarisations at frequencies of ~750 cm$^{-1}$, ~850-900 cm$^{-1}$, ~1100-1200 cm$^{-1}$ and ~3100 cm$^{-1}$, which are independent of the ribbon width for all of our studied nanoribbons.

\subsubsection{Fundamental modes}
In any nanoribbon, a group of six modes can be found that are
equivalent to the six $\Gamma$-point phonon modes of graphene with
respect to the phonon eigenvectors. The two in-plane optical modes, in contrast to the
graphene optical modes, are not found to be degenerate: the in-plane
transverse optical mode (TO) has a higher frequency than the inplane longitudinal optical mode
(LO) for each of our studied AGNRs.

\begin{figure}
\includegraphics[width=0.75\columnwidth]{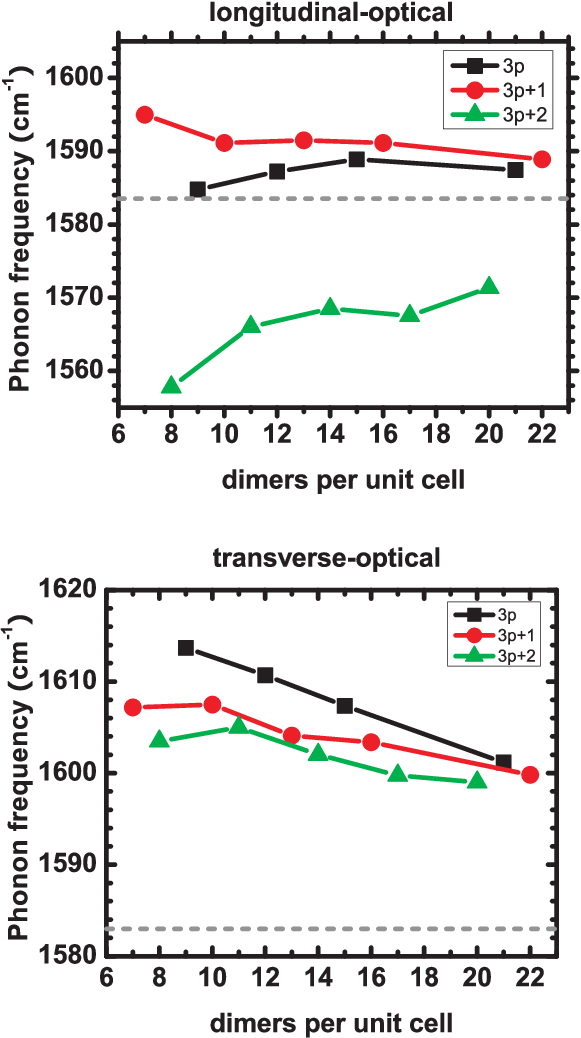}
\caption{\label{fig:FM} LO and TO fundamental mode frequencies of armchair nanoribbons. Our calculated phonon frequencies were scaled by a constant factor $C$=0.975 to achieve a better compatibility of calculations and experimental values. The dashed line indicates the experimental $E_{2g}$ frequency of graphene.}
\end{figure}

The frequencies of these modes are displayed in Fig.~\ref{fig:FM},
together with the frequency of the experimental $E_{2g}$ mode in graphene.
According to Son \emph{et al.}\cite{son216803}, the nanoribbons can be classified into families $N=3p$, $N=3p+1$ and $N=3p+2$, with $p$ a positive integer. The LO-TO-splitting found for ribbons of the $N=3p$ family is about 29 cm$^{-1}$  for the the smallest investigated nanoribbon and about 14 cm$^{-1}$ for the largest one. For the $N=3p+1$ family, we found a splitting of 12-14 cm$^{-1}$ for all investigated nanoribbons. The $N=3p+2$ family displays a larger splitting. It is about 46 cm$^{-1}$ for the 8-AGNR and decreases with increasing ribbon width to a value of about 27 cm$^{-1}$ for a 20-AGNR. All these LO-TO-splittings should be experimentally measureable.
For the LO-modes with $N=3p$ and $N=3p+1$ and the TO-modes an increase
of the frequency compared to graphene is found.
As can be seen, the LO frequencies of the ($3p+2$)-nanoribbons are
softened. This can be attributed to the small band gap in the quasi-metallic $(3p+2)$ nanoribbons, which is smaller than 0.294\,eV for $p>3$. This is similar to the LO phonon softening in metallic carbon
nanotubes\cite{Dubay02}\cite{piscanec07}. We assume that the same effect of strong electron-phonon coupling related to a Kohn anomaly takes place in quasi-metallic GNR\cite{piscanec04}. All modes converge towards the graphene frequency with increasing width.

\subsubsection{Overtones}
For each fundamental oscillation, we find $(N-1)$ overtones, where the
fundamental displacement pattern is modified by an envelope forming a
standing wave with $x=1..N$ nodes. The vibrational behavior of these modes shows similarities to elastic sheets with free ends. The atomic displacement can be described by an enveloping cosine function
\begin{equation*}
f_n = A_n\cos k_{\perp,n}x = A_n\cos \frac{\pi}{\lambda_{\perp,n}}x,
\end{equation*}
where $n$ is the order of vibration, $k_{\perp,n}$, $\lambda_{\perp,n}$ and $A_n$ refer to the wavenumber, the wavelength and the amplitude of the $n$th order vibration. For the wavenumbers hold the following relations.
\begin{eqnarray}
k_{\perp,0} &=& 0\nonumber\\
k_{\perp,n} &=& \frac{2\pi}{\lambda_{\perp,n}}\label{G1}\\
&=& \frac{n \pi}{w_{\mbox{\tiny AGNR}}}\nonumber\\
&=&\frac{2n\pi}{\left(N-1\right)a_0}\label{G3}
\end{eqnarray}
\begin{figure}
\includegraphics[width=\columnwidth]{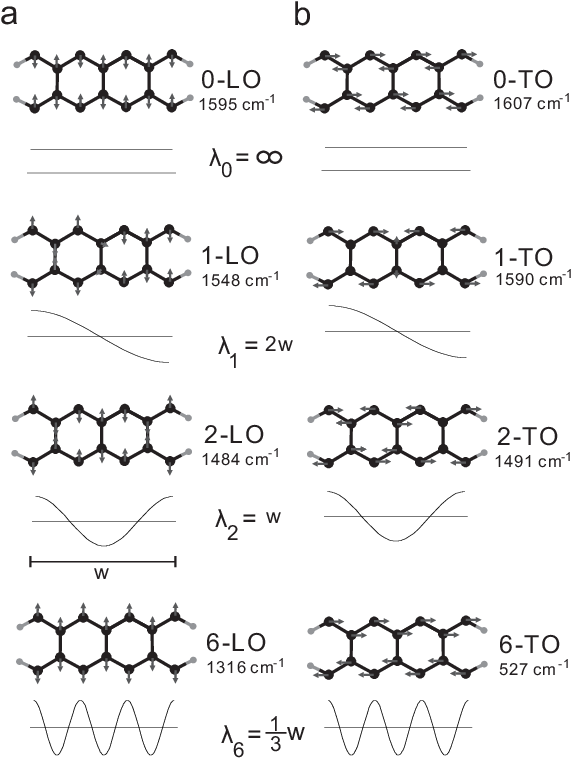}
\caption{\label{fig:LORMTORM} (a) longitudinal-optical (LO) and (b)
  transversal-optical (TO) fundamental and overtone modes at the
  $\Gamma$ point of a 7-AGNR. The arrows display the displacements of
  the atoms in the unit cell. The displacement strength is normalized to emphasize the node positions. For the $n$-L/TO, the eigenvectors of
  the atoms reverse $n$-times 0-L/TO across the ribbon
  width. This is further clarified by the envelope curves. The
  wavelength of the vibrations is $\lambda=\frac{2}{n} w_{\mbox{\tiny AGNR}}$.}
\end{figure}
The nodes do not have to coincide with carbon atom positions in the
unit cell.
Figure~\ref{fig:LORMTORM} shows the displacement patterns of a 7-AGNR.
We characterize the phonon modes by their direction of vibration
(transverse, longitudinal, out-of-plane) and their nature
(acoustic, optical) as $n$-L/T/ZA and $n$-L/T/ZO, with $n$ = number
of nodes.

\begin{figure}
\includegraphics[width=\columnwidth]{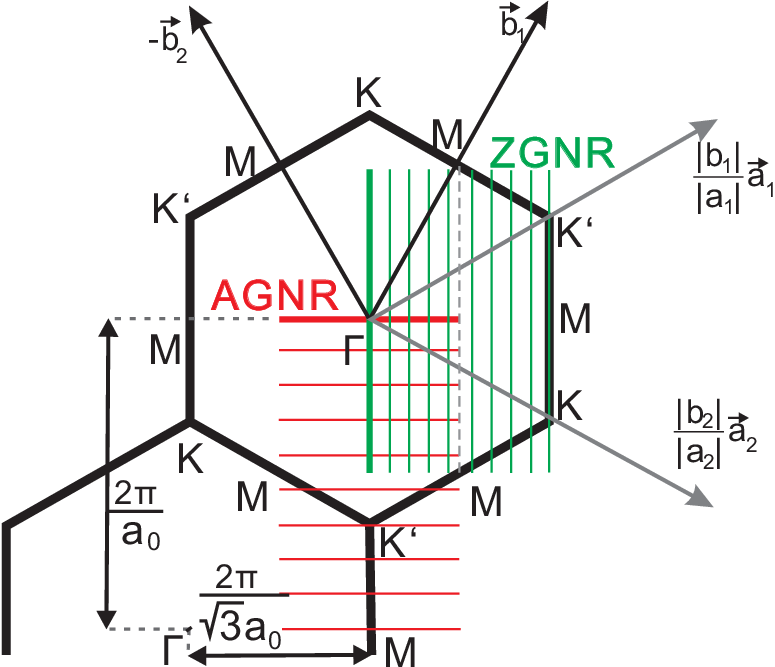}
\caption{\label{fig:brillouin} (color online) Brillouin zones of graphene, armchair (10-AGNR) and zig-zag (10-ZGNR) nanoribbons. $\vec a_1$ and $\vec a_2$ are the lattice vectors of graphene, $\vec b_1$ and $\vec b_2$ are the reciprocal lattice vectors. Note that the Brillouin zone of ZGNRs ist idelaized; in actual nanoribbons it reaches the $K$-$M$-$K'$ line only in the limit of large nanoribbons, see discussion in sect. III C.}
\end{figure}

\subsection{AGNR modes in relation to graphene}
\begin{figure}
\includegraphics[width=\columnwidth]{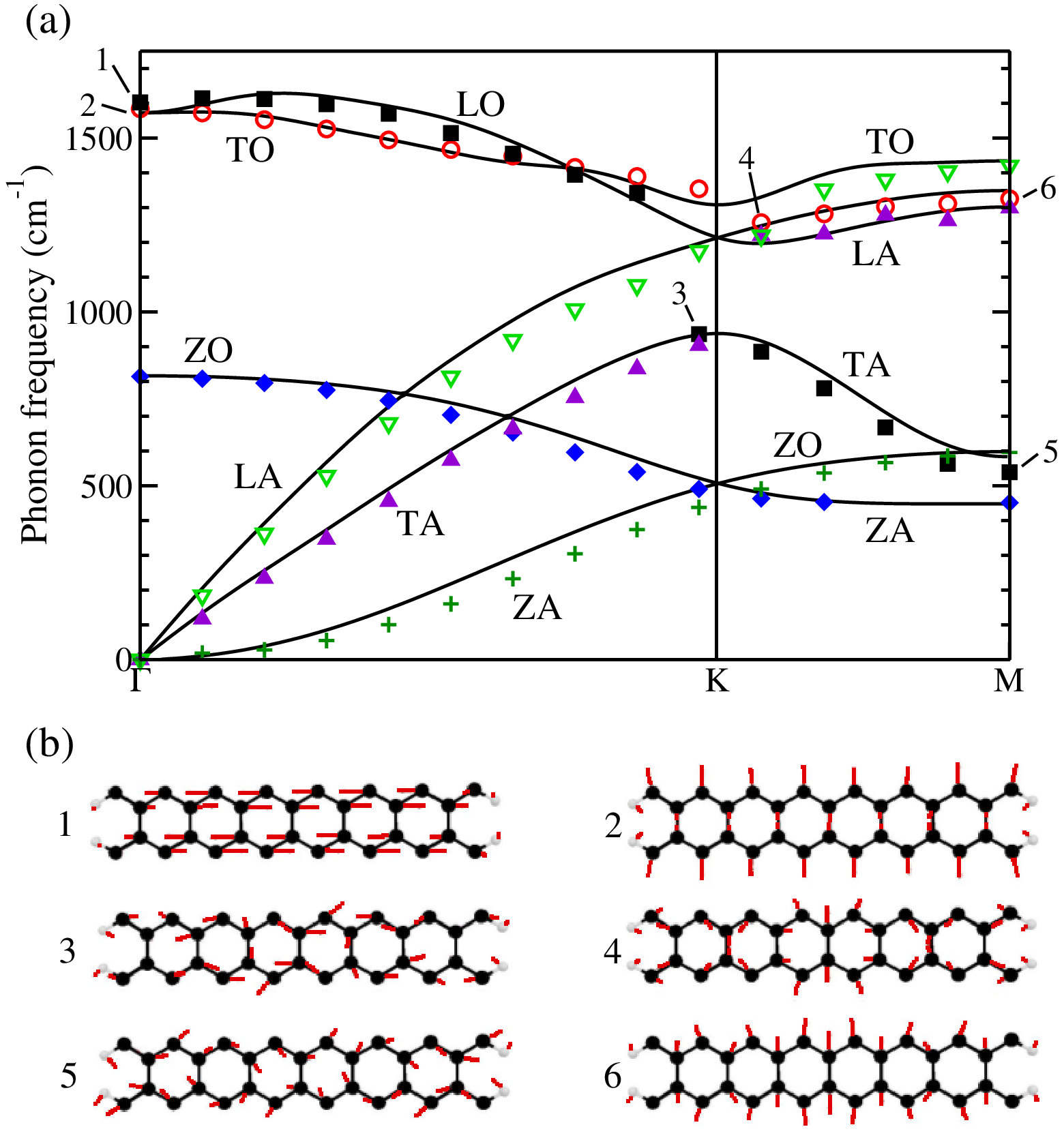}
\caption{\label{fig:baender}(color online) (a) Mapping of TO (filled squares), LO
(circles), ZO (filled diamonds), TA (open triangles), LA (filled triangles), ZA (pluses) fundamental and overtone frequencies of an
15-AGNR onto our calculated phonon dispersion of graphene (solid lines). 
Longitudinal ribbon modes correspond to transverse
graphene modes. (b) Eigenvectors of selected ribbon TO and LO
frequencies in $\Gamma KM$ direction. The corresponding modes are indicated in (a). }
\end{figure}

Figure~\ref{fig:brillouin} shows the Brillouin zones of graphene,
armchair nanoribbons and zig-zag nanoribbons. The hexagonal structure
with high symmetry points $K$ and $M$ represents the Brillouin zone of
graphene with the distances $\overline{\Gamma K}$=${4\pi}/{3a_0}$
and $\overline{\Gamma M}$=${2\pi}/{\sqrt{3}a_0}$.  As already
discussed, the phonon vectors in nanoribbons are restricted by an edge-induced boundary condition, resulting in $N$ quantized wave numbers $k_{\perp,n}$ along the ribbon width (Eq.~\ref{G3}). The component in axial direction however is unrestricted and not quantized. We find that the Brillouin zone of graphene nanoribbons consists of $N$ equally spaced discrete lines, similar to the Brillouin zone of carbon nanotubes\cite{reich04buch}. The line spacing for armchair nanoribbons is, from Eq.~\ref{G3}, $\Delta k_{\perp,n}=\frac{2\pi}{\left(N-1\right)a_0}$. The translation vector of an
armchair nanoribbon is given by $\vec a_{\mbox{\tiny AGNR}}=\vec a_1 + \vec a_2$, i.e. the axial direction of armchair nanoribbons corresponds to the
$\Gamma$M-direction in graphene. The direction perpendicular to the
ribbon axis corresponds to the $\Gamma$KM-direction. The AGNR $\Gamma$-point overtone vibrations therefore correspond to vibrations in $\Gamma KM$ direction. It should be possible to ``unfold'' the Brillouin zone of a nanoribbon onto that of a graphene sheet, where the
$\Gamma$-point frequencies of fundamental and overtone modes reproduce
discrete graphene modes along the $\Gamma KM$ direction. For the overtone of the highest order, i.e. $n=N-1$, we find from Eq.~\ref{G3}
\begin{eqnarray*}
k_{\perp,N-1} &=& \frac{2(N-1)\pi}{(N-1)a_0}\\
&=& \frac{2\pi}{a_0}
\end{eqnarray*}
As can be seen in Fig.~\ref{fig:brillouin}, $|\overline{\Gamma KM}|={2\pi}/{a_0}$, thus it should be possible to reproduce the whole $\Gamma KM$-dispersion of graphene by nanoribbon $\Gamma$-point overtones. Figure ~\ref{fig:baender} shows a mapping of the resulting pairs ($k_{\perp,n}$,$\omega_{n}$) of AGNR $\Gamma$-point phonon modes onto the phonon dispersion of graphene. Our calculated dispersion is in very good agreement with experimentally obtained results\cite{mohr07gr}\cite{maultzsch04}. 
As we defined the longitudinal and transverse direction with respect to the ribbon axis, the $n$th overtone of a longitudinal ribbon mode corresponds to a transverse phonon at wavevector $k_{\perp,n}$ in graphene and vice versa.

The overtones of one respective ribbon fundamental mode
reproduce different phonon branches of graphene in $\Gamma K$- and
$KM$-direction. For example, the ribbon TO frequencies reproduce the
graphene LO branch in $\Gamma K$-direction, but then switch to 
the graphene TA-branch beyond the $K$-point. 
The reason for this lies in the strict mode classification applied to the nanoribbons.
We can also understand the branch switching from the Brillouin zone of graphene (see Fig.~\ref{fig:brillouin}): going along $\overline{\Gamma K}$ and then along $\overline{KM}$, the direction of the wave vector changes by 120$^{\circ}$, if one continues to stay in  the 1$^{st}$ Brillouin zone of graphene. Moreover, it is well-known that close to the K-point the phonon modes loose their purely longitudinal or transverse character.
Figure ~\ref{fig:baender} (b) shows the phonon eigenvectors of different overtones of the nanoribbon modes. It is clearly seen that the overtones near the graphene $K$-point have a mixed character with displacements in different directions [see panels 3 and 4 in Fig.~ \ref{fig:baender} (b)]. In total, only small deviations between the calculated graphene dispersion and the zonefolded nanoribbon frequencies are found. It is expected that ribbon 
frequencies converge towards the graphene dispersion with increasing
ribbon width, a result which we find confirmed. The root mean square deviation
between graphene LO branch and their corresponding ribbon modes
decreases from $59$ cm$^{-1}$ (7-AGNR) to  $31.5$  cm$^{-1}$ (14-AGNR).

\begin{figure}
\includegraphics[width=\columnwidth]{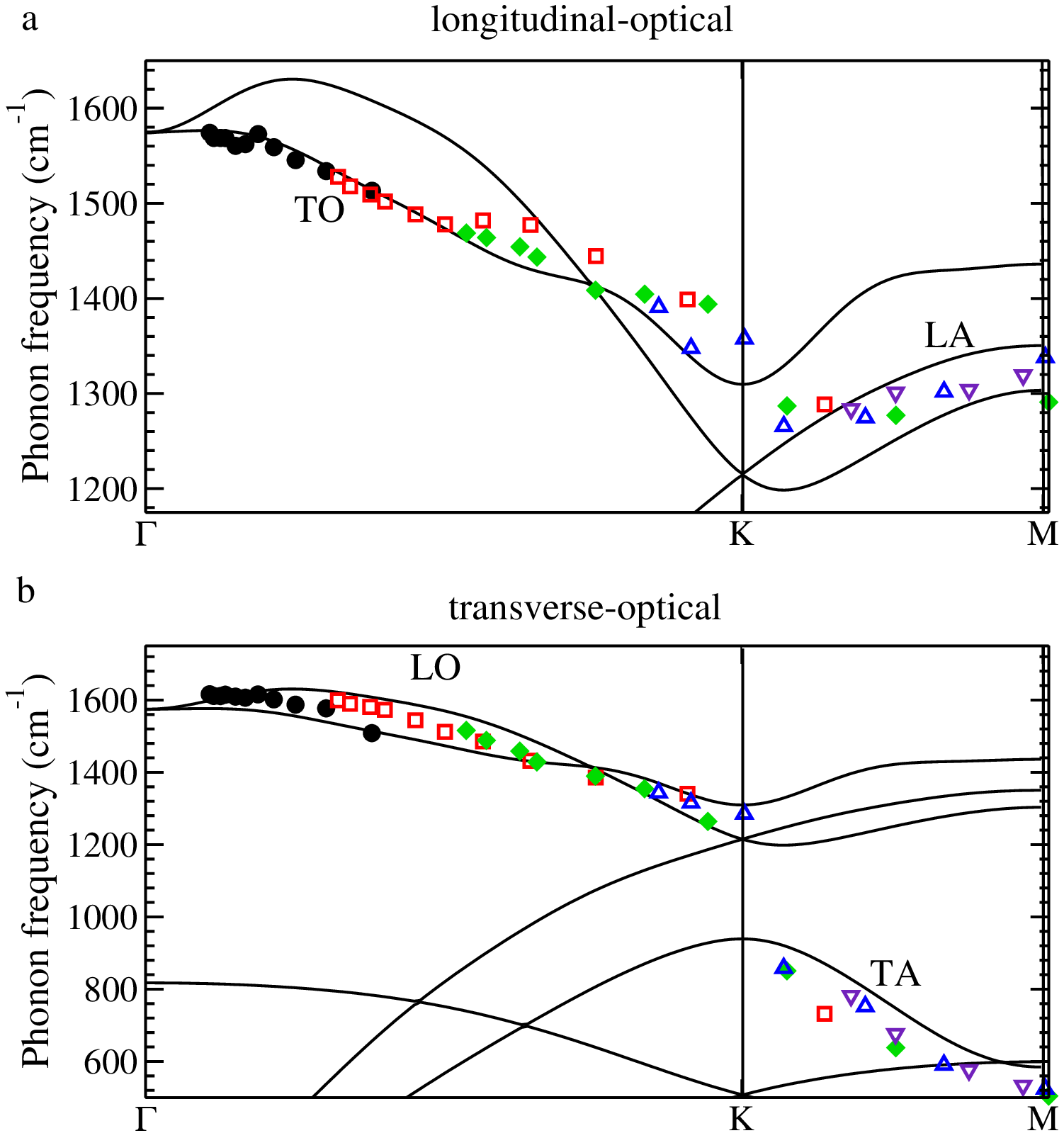}
\caption{\label{fig:LOTO}(color online) (a) $n$-LO and (b) $n$-TO overtones of $N$-AGNRs with $N=5-15$ and $n=1$ (filled circles), $n=3$ (empty squares), $n=5$ (filled diamonds), $n=8$ (empty triangles) and $n=11$ (crosses). Solid black lines are the calculated graphene modes as in Fig.~\ref{fig:baender}.}
\end{figure}

In general, the overtones reproduce the graphene phonon dispersion
fairly well. However, they cannot mimic the special circumstances near
the $K$-point. There, the TO-mode of graphene displays a noticeable drop in frequency due the a Kohn anomaly\cite{pisanibornoppen,piscanec04,maultzsch04}, i.e. a strong electron-phonon coupling. As visible in Fig.~\ref{fig:LOTO} (a), the nanoribbon overtones show a poorer reproduction of this drop of frequency in the surrounding of the graphene $K$-point. This is understandable 
because of the semiconducting nature of the investigated nanoribbons, which would prevent the formation of Kohn anomalies. In this case, armchair nanoribbons with smaller band gaps should reproduce the graphene dispersion near the $K$-point better than AGNRs with larger gaps. Indeed, our calculations suggest that the quasi-metallic AGNRs of the $N=3p+2$-family, which have very small band gaps, give slightly better results than the ribbons of the other families.
As mentioned above, the nanoribbon modes change their character from longitudinal to transverse direction and vice versa when crossing the $K$-point.
In particular, the displacement vectors of the ribbon LO between
$K$ and $M$ show strong similarities with the LA branch in graphene. Correspondingly, the frequencies agree well with the LA branch (Fig.~\ref{fig:LOTO}\,a). Similarly,
the nanoribbon TO modes switch in characteristics to TA eigenvectors when crossing the $K$-point in $M$-direction (Fig.~\ref{fig:LOTO}\,b). We find thus that the overtones of the nanoribbons can be well mapped onto the graphene dispersion when including the character of their eigenvectors. In graphene, the overtones correspond to different branches in the $\Gamma$-$K$ and $K$-$M$ parts of the Brillouin zone.
Overall we encountered difficulties in characterizing modes close to the graphene $K$-point. The characteristic overbending of the graphene LO-mode due to a Kohn anomaly at $\Gamma$-point is found by zone-folding of the ribbon modes, too, for nanoribbons of sufficient width ($N>8$). However, the observed overbending is considerably smaller than the one in the calculated graphene dispersion.

\subsection{Zigzag nanoribbons}
Recent studies show that the ground state of zigzag nanoribbons displays antiferromagnetically ordered spin states\cite{fujita96,sonmetall,PhysRevLett.87.146803}. Calculations using spin-polarization predict the opening of a band gap for the otherwise metallic ZGNRs and half-metallic behavior when an electric field is applied due to the opposite behavior of different spin directions in electric fields\cite{sonmetall}.
We analyzed the phonon spectra of ZGNRs for effects due to the bandgap opening by generating a pseudo potential including spin polarization with using exactly the same cutoff radii as the pseudo potentials we used for calculations neglecting spin-polarization. We find excellent agreement with the results of Son \emph{et al.}\cite{sonmetall} regarding the band gap. While the spin-polarization effects are vital for the electronic properties, we observe only small effects on the vibrational frequencies. Our calculated $\Gamma$-point frequencies differ by just up to $8$ cm$^{-1}$  from calculations neglecting spin polarization. For this reason, we did not include spin-polarization in the following phonon calculations.

\begin{figure}
\includegraphics[width=0.9\columnwidth]{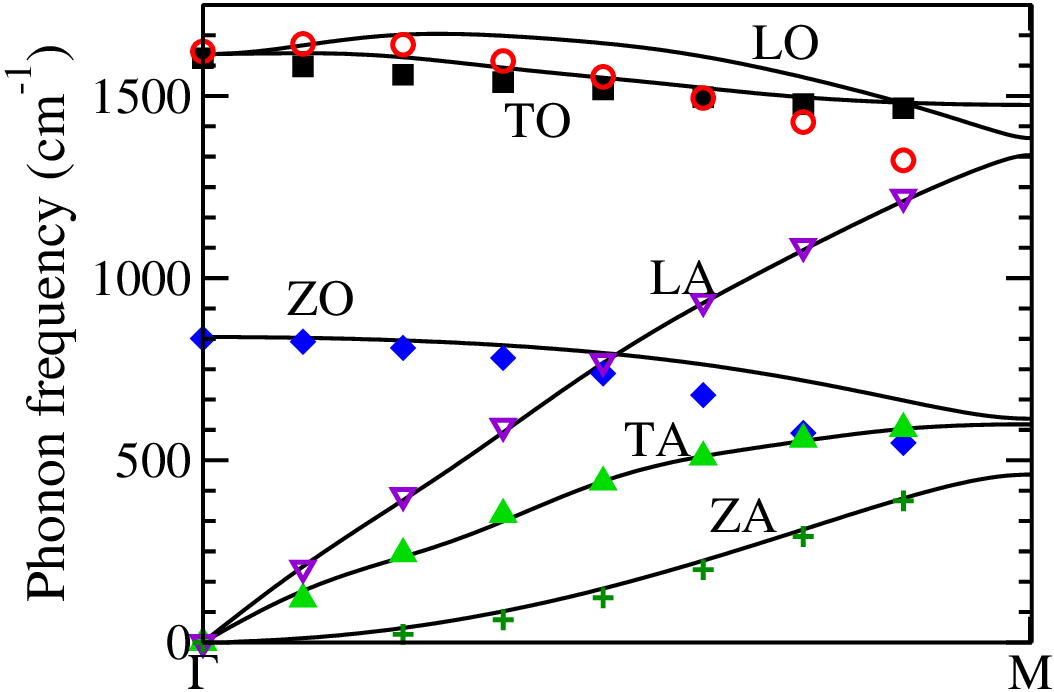}
\caption{\label{fig:ZGNRdisp} (color online) Mapping of ribbon LO (filled squares), TO (circles), ZO (filled diamonds), TA (down triangles), LA (filled up triangles), ZA (crosses) fundamental and overtone frequencies of an 8-ZGNR onto the calculated phonon dispersion of graphene (solid lines). $k_{\perp, N-1}$ does not reach the graphene $M$-point, as for ZGNRs of small width, the wavelengths obtained by fitting the displacement pattern of the calculated phonons are larger than the theoretically predicted ideal values.}
\end{figure}

In order to study the vibrational behavior of zigzag nanoribbons, we carried out calculations on $N$-ZGNRs with $N=2..14$ and performed the same rescaling of the calculated frequencies as was done for AGNRs. A distinction of the $\Gamma$-point phonons in fundamental modes, overtones and C-H-modes is performed as for the armchair nanoribbons. The direction perpendicular to the ribbon axis reproduced by the mapping corresponds to the $\Gamma$M-direction of graphene, see Fig.~\ref{fig:brillouin}. The wavelengths of the vibrations of equivalent carbon atoms over the nanoribbon width can be described by
\begin{equation}
\lambda_n = \frac{2}{n}w_{\mbox{\tiny ZGNR}}\label{G6}\\
\end{equation}
By Eqs.~\ref{G5},~\ref{G1} and~\ref{G6}, we calculate a line spacing of 
\begin{eqnarray*}
\Delta k_{\perp} &=& \frac{2\pi}{\lambda_{n+1}}-\frac{2\pi}{\lambda_{n}}\\
&=& \frac{2\pi}{\sqrt{3}(N-1)a_0}\\
\end{eqnarray*}
The overtone of highest order, $k_{\perp,\mbox{\tiny N-1}}$, is then
\begin{eqnarray*}
k_{\perp,\mbox{\tiny N-1}} &=& \Delta k_{\perp}(N-1)\\
&=& \frac{2\pi}{\sqrt{3}(N-1)a_0}(N-1)\\
k_{\perp,\mbox{\tiny N-1}} &=& \frac{2\pi}{\sqrt{3}a_0}\\
\end{eqnarray*}
This is equal to the graphene $\Gamma$-M distance, as $|\overline{\Gamma M}|=\frac{2\pi}{\sqrt{3}a_0}$.
Therefore, we can, in theory, reproduce the whole $\overline{\Gamma M}$ of the graphene dispersion by unfolding the $\Gamma$-point phonons of ZGNRs of finite  width. On the other hand, we determined the wavelengths of the overtones of our investigated ribbons by fitting a cosine function to the respective displacement patterns and compared them to the theoretically expected values. For small nanoribbons, we found considerable deviations between the wavelength of the lattice vibration of graphene at the M-point and the smallest wavelength that the atomic displacements of the nanoribbon can describe, i.e. the wavelengths of the highest order overtones. Thus, the mapping of the Brillouin zone of small nanoribbons cannot reproduce the whole $\Gamma M$-direction, as $k_{\perp,\mbox{\tiny  N-1}}<k_{\mbox{\tiny M-point}}=\frac{2\pi}{\sqrt{3}a_0}$. However, these smallest wavelengths quickly converge towards the graphene M-point wavelength with increasing ribbon width. 
We performed mappings of the phonon modes of ZGNRs of various widths onto the graphene phonon dispersion in $\Gamma M$-direction (Fig.~\ref{fig:ZGNRdisp}).
Again, unfolding the ribbon overtones onto the Brillouin zone of graphene shows good agreement, which improves for increasing ribbon width. The overtones of highest order of the optical modes of small nanoribbons appear to be clinched due to the mentioned deviations of fitted and theoretically predicted wavelength. The acoustical modes, however, display a great agreement of nanoribbon overtones and graphene dispersion. The ZGNR fundamental mode frequencies correspond to the six $\Gamma$-point frequencies of graphene. For ZGNRs we see, in contrast to AGNRs, a clear separation between the frequencies of in-plane acoustic and optical phonon modes. The in-plane acoustic modes are found in a frequency interval of 0-1300 cm$^{-1}$, the (in-plane) optical modes lie between 1300-1600 cm$^{-1}$ and
are straightforward to classify as all of them are of pure longitudinal or transverse nature. The displacement pattern of ZGNRs shows no apparent mixing of modes, as found for AGNRs. As mentioned in Sect. III B, the mixing of modes in AGNRs occurs due to the symmetries at the graphene $K$-point. However, there's no comparable point in the $\Gamma M$-direction, which is reproduced by the mapping of the ZGNR Brillouin zones onto the one of graphene.

\begin{figure}
\includegraphics[width=0.9\columnwidth]{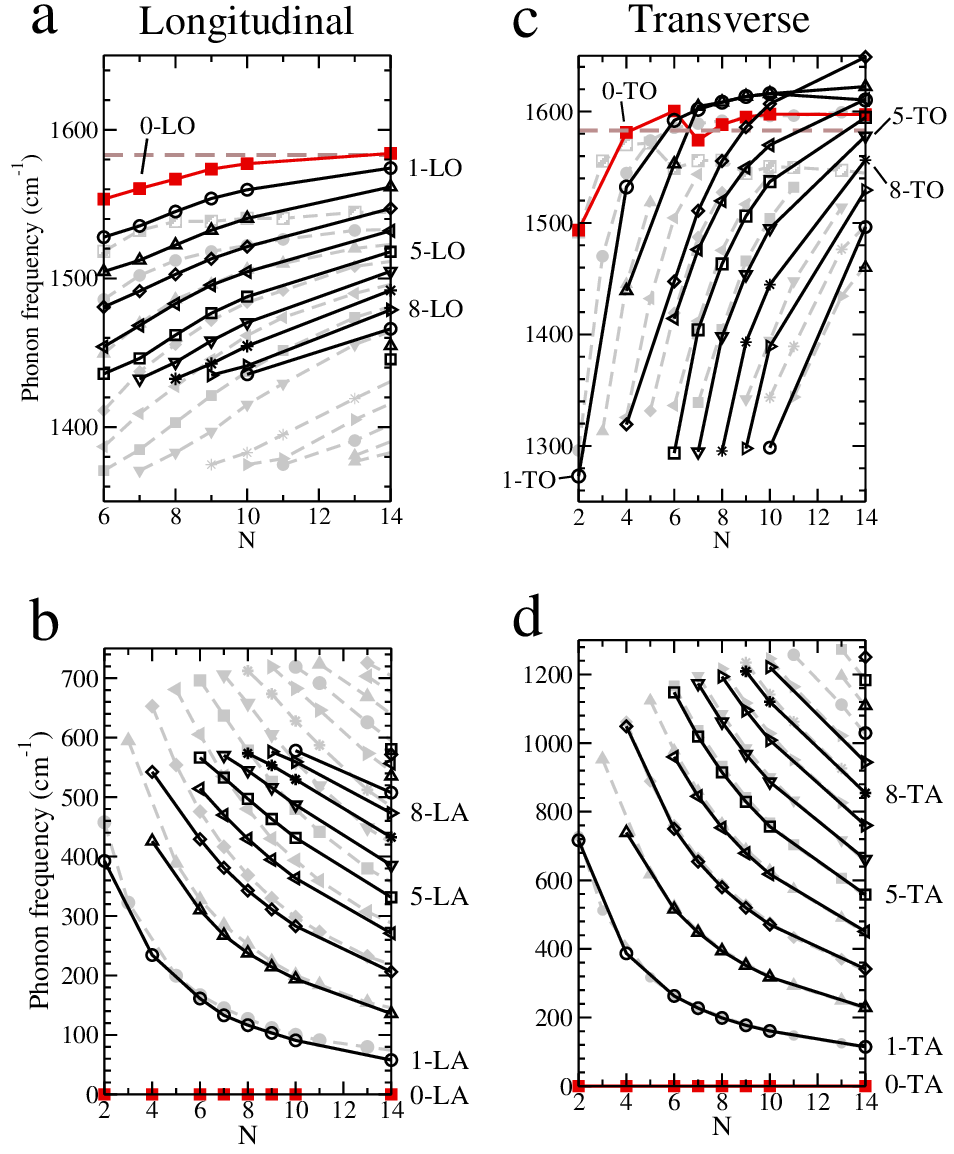}
\caption{\label{fig:ZOS} (color online) Calculated longitudinal (a,b) and transverse (c,d) $\Gamma$-point frequencies of the ZGNR in dependence of nanoribbon width. Filled (red) squares are fundamental oscillations, empty symbols are overtones. Solid black lines connect overtones of equal order $n$ for different ribbons.
Filled grey symbols connected with dashed lines show the results of calculations of Yamada \emph{et al.}\cite{yamada08} for comparison. The experimentally determined $E_{\mbox{\tiny 2g}}$ mode frequency of ~1580 cm$^{-1}$ is indicated by thick (brown) dashed lines.}
\end{figure}

Figure ~\ref{fig:ZOS} shows a comparison of in-plane phonon mode frequencies of $N$-ZGNRs with $N=2..14$. As was found for AGNRs, the 0-LO of the ribbon converges towards the graphene LO for increasing ribbon width. A similar behavior is found for the longitudinal optical overtones. For the frequency of the 0-TO, a non monotonic dependence on the ribbon width is observed. The calculated frequencies of the transverse optical overtones of low order are higher than those of the 0-TO. Similarly as for the AGNRs we thus find for ZGNR a (small) overbending for the graphene LO mode with our zone-folding method, at least for sufficiently large ribbon widths. The acoustic overtones of both longitudinal and transverse nature display an inversely proportional width dependence. In case of transverse overtones, this width dependence is well described by $\omega_{ac}\propto N^{-1}$. Longitudinal acoustic overtones, however, show a considerably weakening width dependence with increasing vibrational order, as can be seen in Fig.~\ref{fig:ZOS} (b).

\begin{figure}
\includegraphics[width=\columnwidth]{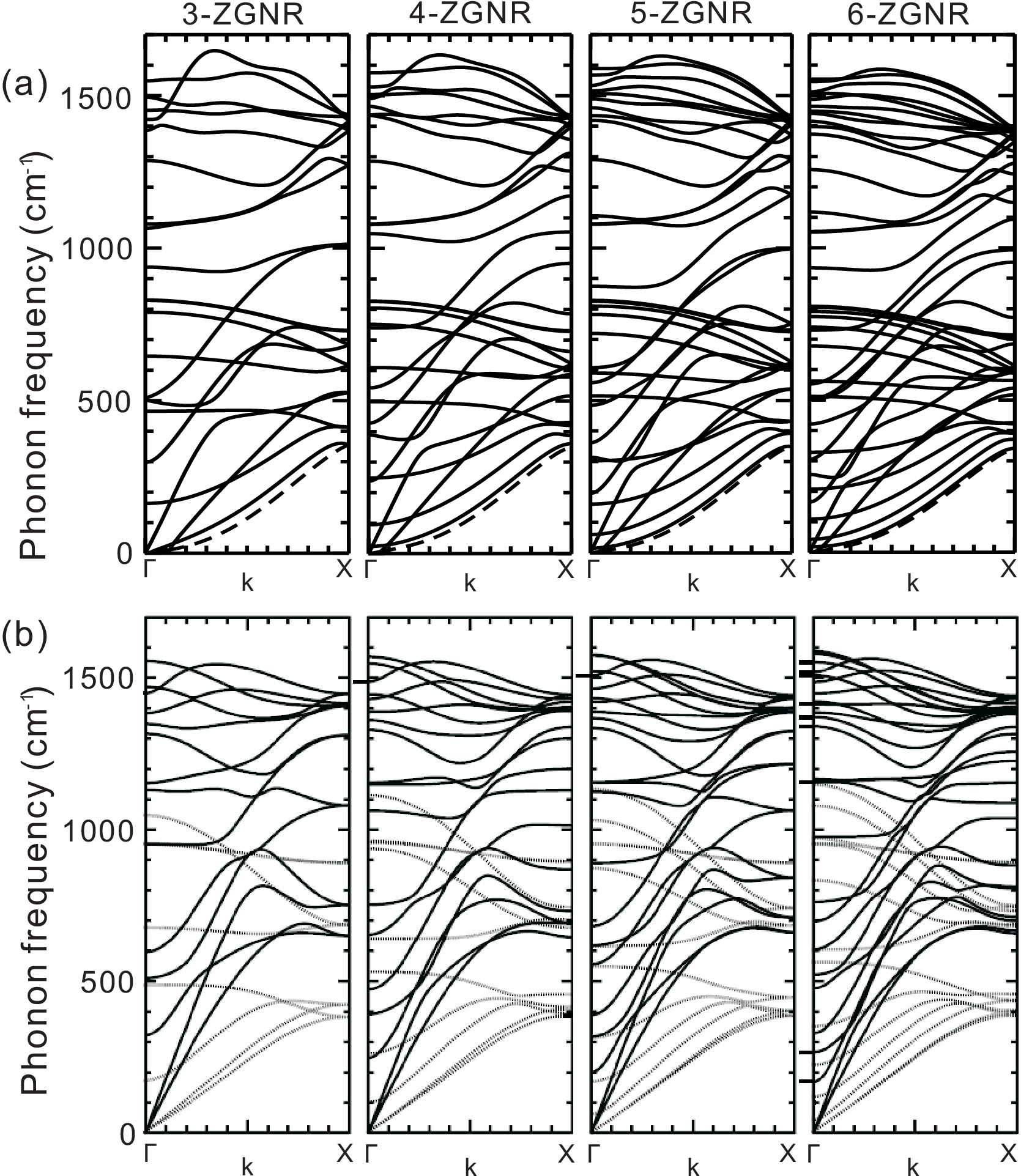}
\caption{\label{fig:dispersions} Phonon dispersions of hydrogenated $N$-ZGNRs with $N$=2-6 obtained by (a) DFT and (b) MO/8 calculations done by Yamada \emph{et al}\cite{yamada08}. The dashed lines in (a) indicate the fourth acoustic mode, typical for 1D-crystals. Dashed lines in (b) are out-of-plane vibrations.}
\end{figure}

Finally, we calculated the phonon dispersions over the whole Brillouin
zone of various small nanoribbons by means of a supercell
approach. We used a supercell of 9 unit cells along the nanoribbon axis. Figure ~\ref{fig:dispersions} (a) shows the dispersions of
$N$-ZGNR with $N=3-6$. As can be seen, the dispersions feature the characteristic fourth acoustic mode of 1D
structures, which is a rotational mode around the z-axis. The displacement pattern of this mode
at the $\Gamma$-point corresponds to the displacement
pattern of the mode we classify as $1$-ZA. However, the $1$-ZA in our calculations has a frequency
$\omega_{\mbox{\tiny 1ZA}}$=$5-20$ cm$^{-1}$, which we believe results from the presence of the hydrogen passivation and possibly also from numerical errors. The two out-of-plane acoustic modes converge swiftly
for increasing ribbon widths, being noticeably
separated for the $3$-ZGNR, but almost degenerate for the $6$-ZGNR,
closely resembling the ZA-mode of graphene. 
An interesting fact was found for the phonon modes at the $X$-point: In armchair nanotubes the phonon modes are pairwise degenerate, i.e 6(n-1) phonon modes in $(n,n)$-CNTs with odd $n$ and all modes in $(n,n)$-CNTs with even $n$, at the $X$-point due to symmetry\cite{reich04buch}. Therefore, we might exüect $6(N-1)$ pairwise degenerate and 6 non-degenerate modes in $N$-ZGNR with odd $N$.
Similarly, all phonon modes of $N$-ZGNRs with even $N$ should be pairwise degenerate. However, we do not find a similar degeneracy for the zigzag nanoribbon dispersions we studied so far. In fact, the calculated phonon spectra of $N$-ZGNR with odd $N$ consist solely of modes that are pairwise degenerate at the $X$-point. In contrast, the phonon modes of $N$-ZGNR with even $N$ are largely non-degenerate at the edge of the Brillouin zone.

We want to compare our results with previous calculations by Yamada \emph{et al.}\cite{yamada08}. They applied the MO/8-method\cite{ohno02}, which uses force fields based on the Hückel molecular orbital theory, i.e. a semi-empirical approach. This approach was shown to be efficient for calculating the vibrational properties of polycyclic aromatic hydrocarbons like graphene. In this approach, the topology of the structures of interest is fixed consisting of hexagons with C-C bond lengths of 1.39 \AA\space and C-H bond lengths of 1.048 \AA. As can be seen in Fig.~\ref{fig:ZOS} and ~\ref{fig:dispersions} (b),
we find general agreement, but some deviations in particular for the
longitudinal modes. We suggest the geometry relaxation that we performed in our calculations to be responsible for these deviations.

\section{Conclusion}
We investigated the vibrational properties of graphene nanoribbons 
with density functional theory.  We showed that the
$\Gamma$-point phonons of graphene nanoribbons with armchair and
zigzag type edges can be interpreted as six fundamental oscillations
and their overtones, which show a characteristic nanoribbon width
dependence. We demonstrated that the $\Gamma$-point phonon frequencies
of nanoribbons can be mapped onto the phonon dispersion of graphene,
i.e. to an "unfolding" of the nanoribbons' Brillouin zone onto that of
graphene. The edge magnetization and the resulting opening of a band
gap in zig-zag nanoribbons has only a small influence on the phonon
spectra. The behavior of overtones and fundamental modes for
nanoribbons of increasing width was studied and a comparison of our
results for ZGNRs with past studies performed.

\section{Acknowledgements}
This work was supported in part by the Cluster of
Excellence 'Unifying Concepts in Catalysis' coordinated by the TU Berlin and
funded by DFG.

\end{document}